\documentclass[%
 aip,
 amsmath,amssymb,
reprint,%
]{revtex4-1}

\usepackage{graphicx}% Include figure files
\usepackage{dcolumn}% Align table columns on decimal point
\usepackage{bm}% bold math
\usepackage[utf8]{inputenc}
\usepackage[T1]{fontenc}
\usepackage{mathptmx}
\usepackage{etoolbox}
\usepackage{xcolor}
\usepackage{siunitx}
\usepackage[version=4]{mhchem}

\makeatletter
\def\@email#1#2{%
 \endgroup
 \patchcmd{\titleblock@produce}
  {\frontmatter@RRAPformat}
  {\frontmatter@RRAPformat{\produce@RRAP{*#1\href{mailto:#2}{#2}}}\frontmatter@RRAPformat}
  {}{}
}%
\makeatother

\begin{document}

%\preprint{AIP/123-QED}

\title{Dielectric Tensor of CrSBr from Spectroscopic Imaging Ellipsometry}

% Force line breaks with \\
\author{Pierre-Maurice Piel}
%\email{ppiel@uni-muenster.de}
\thanks{equal contribution}
 \affiliation{Institute of Physics, University of M\"unster, M\"unster, Germany}
\author{Sebastian Schaper (n\'e Funke)}
\thanks{equal contribution}
\affiliation{Institute of Physics, University of M\"unster, M\"unster, Germany}
\author{Aleksandra Łopion}
\affiliation{Institute of Physics, University of M\"unster, M\"unster, Germany}
\author{Jakob Henz}
\affiliation{Institute of Physics, University of M\"unster, M\"unster, Germany}
\author{Aljoscha Soll}
\affiliation{Department of Inorganic Chemistry, University of Chemistry and Technology Prague, Prague, Czech Republic}
\author{Zdenek Sofer}
\affiliation{Department of Inorganic Chemistry, University of Chemistry and Technology Prague, Prague, Czech Republic}
\author{Ursula Wurstbauer}
\affiliation{Institute of Physics, University of M\"unster, M\"unster, Germany}
\affiliation{Center for Soft Nanoscience (SoN), University of Münster, Germany}
\email{wurstbauer@uni-muenster.de}

\date{\today}% It is always \today, today,

\begin{abstract}
Chromium sulfur bromide (CrSBr) is a magnetic van der Waals semiconductor with a direct bandgap and pronounced anisotropy in its electronic, optical, spin and lattice degrees of freedom. Here, we employ spectroscopic imaging ellipsometry (SIE) and Mueller-matrix analysis to determine the full dielectric tensor of paramagnetic CrSBr thin films. Our measurements reveal optical anisotropy, characterized by three distinct diagonal components of the dielectric tensor. The in-plane elements are dominated by prominent excitonic resonances polarized along the two main crystallographic axes. Two main excitonic bands (A and B excitons) centered around 1.3 eV and 1.7 eV,respectively, are identified; the A-exciton polarized along the $b$-crystallographic direction, whereas the B-exciton appears to consist of two nearly degenerate contributions polarized along two orthogonal in-plane crystal axes. These results provide fundamental insight into anisotropic light–matter interactions in CrSBr, relevant for future spin-optoelectronic and photonic applications.
\end{abstract}
\maketitle

%%%%%%%%%%%%%%%%%%%%%%%%%%%%%%%%%%%%%%%%%%%%%%%%%%%%%%%%%%%%%%%%%%
% Intro
%%%%%%%%%%%%%%%%%%%%%%%%%%%%%%%%%%%%%%%%%%%%%%%%%%%%%%%%%%%%%%%%%%

Van der Waals (vdW) materials, particularly 2D semiconductors, have garnered significant attention due to their unique tunable electronic and optical properties. 2D magnetic semiconductors combine those advantageous properties with magnetism making them promising multifunctional materials for next-generation spintronic, quantum and neuromorphic devices \cite{2016_Jungwirth, 2021_Kaspar, 2025_Lopion}. Among magnetic vdW semiconductors, chromium sulfur bromide (CrSBr) stands out owing to its stability in ambient conditions and direct band gap in the near-infrared range, which remains largely independent of the number of layers \cite{2021_Wilson, 2023_Klein, 2024_Ziebel}. The orthorhombic crystal structure gives rise to strong anisotropies in lattice, magnetic, electronic, and optical degrees of freedom \cite{2021_Wilson, 2023_Klein, 2023_Torres, 2022_Dirnberger, 2024_Ziebel}. At room temperature, CrSBr exhibits paramagnetic order \cite{2024_Ziebel}, whereas at low temperatures it displays long-range antiferromagnetic order with triaxial magnetic anisotropy \cite{2021_Yang, 2021_Wilson}. Each individual layer is ferromagnetically ordered, whereas the magnetic moments of adjacent layers are anti-parallel aligned with critical Curie and Neél temperatures of T$_C=146$\,K and T$_N$=132\,K, respectively \cite{2021_Yang, 2021_Wilson}. CrSBr's conduction band is nearly flat along $\Gamma - X$-direction in momentum space \cite{2021_Wilson, 2023_Klein, 2025_Heißenbuettel}, resulting in quasi-one-dimensional electronic states and linearly polarized excitons, which are Coulomb-bound electron-hole pairs with cigar-shaped wave functions due to this anisotropy \cite{2023_Klein, 2025_Heißenbuettel}. A peculiarity of CrSBr is the reported self-hybridization of polaritons in crystals thicker than a few tens of nanometers \cite{2022_Dirnberger} already highlighting the importance and strength of the materials light-matter interaction. In CrSBr, not only do excitons and photons interact strongly, but also other degrees of freedom, such as excitons, magnons, phonons, and other collective excitations \cite{2022_Bae, 2024_Ziebel, 2025_Ziegler, 2026_Dirnberger}.

\begin{figure}[h]
  \centering
  \includegraphics[width=0.45\textwidth]{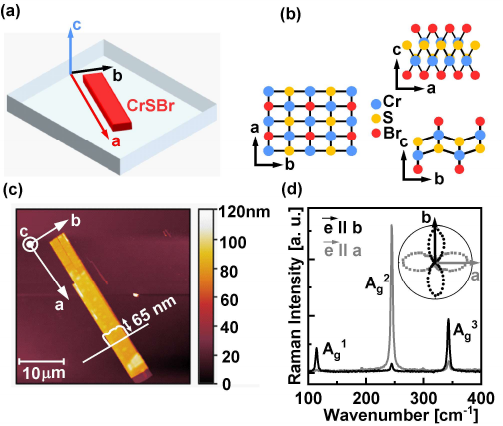}
  \caption{(a) Schematic illustration of the in-plane crystallographic axes $a$ (red) , $b$ (black) and the out-of-plane $c$ axis (blue) for a  CrSBr flake on a substrate. (b) Stick-and-ball model of the orthorhombic crystal structure of CrSBr, showing cuts through the ($a,b$)-, ($c,a$)-, and ($c,b$)-planes. (c) Atomic force microscopy (AFM) topography of a typical CrSBr thin film, with the line profile indicating a thickness of $65 \pm 5$\,nm. (d) Linearly co-polarized Raman spectra of CrSBr ($\vec{e}\parallel a$ and $\vec{e}\parallel b$) at room temperature showing the active modes $A^{\text{1}}_{\text{g}}$, $A^{\text{2}}_{\text{g}}$, and $A^{\text{3}}_{\text{g}}$. The inset displays a polar plot of $A^{\text{2}}_{\text{g}}$ and $A^{\text{3}}_{\text{g}}$ mode intensities with respect to the crystallographic directions.}
    \label{Figure_1}
\end{figure}

The linear optical response of the material is described by the dielectric tensor or equivalently by the refractive index tensor that are both of second rank, where the elements are complex values and dispersive, i.e. a $3\times3$ matrix of complex valued functions \cite{2007_Fujiwara}. Accurate knowledge of the full dielectric tensor is essential for precise modeling of light-matter interaction, prediction and control of excitonic responses, and ultimately guiding the design of next-generation (spin-)optoelectronic devices using CrSBr.\\

\begin{figure}[h]
    \centering
    \includegraphics[width=0.5\textwidth]{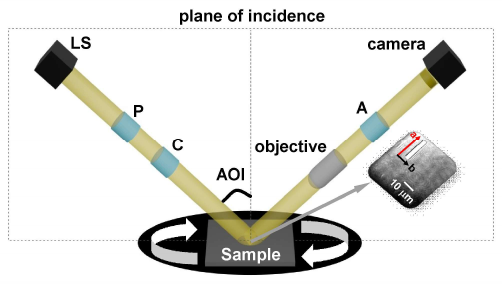}
    \caption{Schematic overview of the spectroscopic imaging ellipsometry (SIE) measurement technique. The change in the polarization state of incident light in reflection from a CrSBr flake placed atop the substrate is measured ina polarizer (P), compensator (c), sample (S) and analyzer (A) geometry (PCSA). Employing a rotation stage enables the Mueller matrix approach. This methodology yields access to the complete dielctric tensor of CrSBr.}
    \label{Figure_2}
\end{figure}

The anisotropic nature of CrSBr does not allow to determine the dielectric tensor with high precision directly from reflectance or absorption experiments as can be done for instance for semiconducting transition metal dichalcogenides that exhibit isotropic response within the 2D plane \cite{2014_Li}.
To overcome this limitation imposed by the anisotropic nature, we employ spectroscopic imaging ellipsometry (SIE) in the visible to near-infrared spectral range \cite{2010_Wurstbauer, 2016_Funke, 2021_Funke_Schiek, 2007_Fujiwara} using both the variable angle Mueller-matrix (MM) approach and generalized ellipsometry (GE) \cite{2021_Funke_Schiek,2007_Fujiwara}. The MM approach is robust against depolarized components of the light reflected from the multilayer stack composed of optical anisotropic layers \cite{2021_Funke_Schiek,2007_Fujiwara,1987_Azzam}. MM allows the experimental determination of the diagonal elements of the dielectric tensor $\varepsilon_{i}(E)$ (\textit{i = a, b, c}) of CrSBr with $\varepsilon_{a}(E) \neq \varepsilon_{b}(E) \neq \varepsilon_{c}(E)$ with high precision and independent from accurate alignment of plane of incidence and crystallographic direction. The in-plane components $\varepsilon_{a}(E)$ and $\varepsilon_{b}(E)$ are dominated by strong excitonic resonances with different excitonic signatures polarized along different crystallographic axes. The principal optical axes and, hence, the diagonal tensor elements correspond to the main crystallographic axes $a, b, c$ as introduced in the schematics in Figure \ref{Figure_1}(a–c). We find good agreement between MM and GE results. For the latter approach, we focus on the in-plane components, keeping $\varepsilon_c(E)$ constant as the function determined via MM. To extract $\varepsilon_a(E)$ and $\varepsilon_b(E)$ the sample is measured using two-zone rotating compensator ellipsometry at two different azimuthal angles. To extract the in plane components, the two datasets are fit to a combined biaxial optical model. The plane of incidence (Figure \ref{Figure_2}) is aligned along the $a$- and $b$-crystal axes, respectively, that are independently determined from polarization resolved Raman spectroscopy (see Figure \ref{Figure_1} (d)). Since we limit the GE approach to the in-plane components, measurements using a fixed angle of incidence (AOI) of 50° are sufficient. Both experiments are carried out on few layer CrSBr samples in the paramagnetic state at room temperature. 

In Figure \ref{Figure_1}(a, b) a schematic of the crystallographic axes $a$ (red), $b$ (black) and $c$-axis (blue) for a CrSBr on a substrate are depicted together with a stick-ball model of the orthorhombic crystal structure. The studied CrSBr thin films are prepared by micromechanical exfoliation from  bulk crystals and transferred onto the substrate by viscoelastic stamping \cite{2014_Gomez}. The topography and height of the thin films are determined by atomic force microscopy (AFM). An AFM topography image of a typical CrSBr flake on a rutile substrate is displayed in Figure \ref{Figure_1}(c) with a thickness of $65 \pm 5$nm.
 
The crystallographic axes are independently determined by polarization resolved Raman spectroscopy at room temperature using an excitation laser of $\lambda$ = 488nm. The linear polarization of the incoming light is rotated with respect to the crystal lattice and the scattered light is determined in co-polarization. As shown in Figure \ref{Figure_1}(d), the Raman active phonon modes A$^{1}_{g}$, A$^{3}_{g}$ are polarized parallel to the $b$-axis and the A$^{2}_{g}$ mode is polarized parallel to the $a$-axis consistent with the anisotropic Raman tensor imposed by the crystal symmetry  \cite{2023_Klein, 2023_Torres}. The strong polarization contrast underscores the pronounced in-plane anisotropy, highlighting the necessity for a full tensorial description of CrSBr. 

\begin{figure*}
    \centering 
    \includegraphics[width=0.9\textwidth]{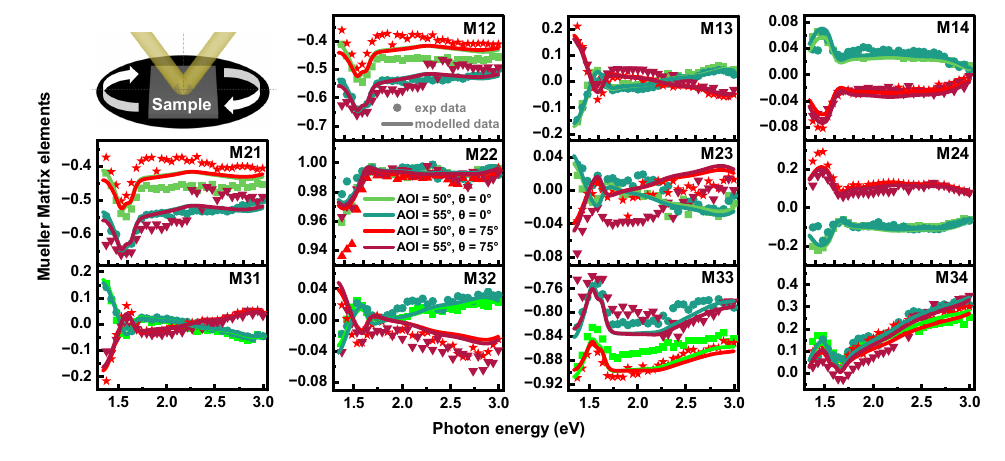}
    \caption{Measured and modeled Mueller-matrix elements $M_{ij}(E)$ for CrSBr thin films on rutile substrates. Mueller-matrix spectra are shown for two angles of incidence (AOI = 50° and 55°) and two sample azimuths ($\theta$ = 0° and 75°), capturing the pronounced optical anisotropy of CrSBr and rutile. Experimental data (symbols) are compared to regression-based optical multilayer modeling (solid lines), demonstrating excellent agreement across all channels. Distinct spectral features corresponding to the A-exciton (~ 1.3\,eV) and B-exciton (~ 1.7\,eV) regimes are clearly resolved in multiple matrix elements, highlighting the axis-dependent excitonic response. The schematic inset illustrates the imaging ellipsometry setup, including the rotating sample stage. [CrSBr on rutile substrate at 300K].}
    \label{Figure_3}
\end{figure*}

To accurately disentangle the dielectric functions and film thickness, the thin exfoliated  CrSBr flakes are transferred onto a transparent birefringent rutile (TiO$_2$) substrate \cite{2024_Schaper, 1951_DeVore, 1997_Jellison} and MM-SIE is initially employed.  By leveraging the out-of-plane component $\varepsilon_c$ determined from the MM approach, generalized SIE is applied to CrSBr placed on isotropic transparent glass (borofloat) substrates and the in-plane components of the dielectric tensor are extracted. 
An ellipsometry measurement quantifies how the polarization state changes upon reflection (or transmission). For the generalized approach, the observables are the amplitude ratio $\Psi$ and phase difference $\Delta$ between $p$ and $s$ components of the reflected light, which are derived from the complex $2 \times 2$ Jones matrix $J$ of the sample considering all contributing layers. Hereby, $\Psi$(E) and $\Delta$(E) are called ellipsometric angles \cite{2007_Fujiwara, 2022_Hilfiker}. While the Jones matrix $J$ connects the Jones vectors of incoming to outgoing light, the Mueller matrix $M$ maps the incoming Stokes vector to the outgoing Stokes vector providing a full description of both the intensity and the state of polarization of the light reflected from the multilayer surface \cite{2007_Fujiwara}. In this way, a complete description of polarization and depolarization components is provided by the Mueller matrix elements $M_{ij}$. As a consequence, the Jones formalism is limited to fully  polarized and coherent light. For anisotropic layer stacks in general, access to the off-diagonal elements $M_{ij}$ of M is critical as $\Psi$ and $\Delta$ alone do not capture the entire optical response of the material \cite{1987_Azzam, 2021_Funke_Schiek, 2007_Fujiwara, 2024_Schaper}.

Figure \ref{Figure_2} shows the SIE geometry in $PCSA$ configuration (linear polarizer $P$, compensator $C$, sample $S$, analyzer $A$) at a chosen angle of incidence (AOI). The sample is globally illuminated with nearly monochromatic light with a well known polarization state. The light reflected from the sample surface is imaged by an objective with an ultra-low numerical aperture (NA < 0.45). The light is then guided through the analyzer and displayed on a CCD detector allowing lateral resolution better than 2$\mu$m. The change of the polarization state is deduced from the configuration of the $P$, $C$ and $A$ elements (for details see SI). By applying a rotation stage under the sample, the Mueller matrix technique can be employed.  With this geometry we have access to two complementary data sets: (i) GE spectra $\Psi(E)$ and $\Delta(E)$ for plane of incidence   $\parallel a$ and $\parallel b$, respectively, for CrSBr placed on an isotropic borosilicate (Borofloat) substrate; and (ii) the $M_{ij}(E)$ elements of the first three lines of the Mueller matrix of the sample that are normalized to its $M_{11}$ element. The $M_{ij}(E)$  are measured for two AOIs (50°, 55°) and two azimuth angles ($\theta$ = 0°, 75°), providing cross-polarization sensitivity and, by the variable AOI, also access to the out-of-plane contribution of the dielectric tensor $\varepsilon_{c}(E)$. The MM elements $M_{ij}(E)$ as well as the ellipsometric angles $\Psi(E)$ and $\Delta(E)$ are experimentally determined in the spectral range covering phonon energies $E$ between 1.2eV and 3eV.   

The dielectric function describing the polarizability of a material exposed to electromagnetic irradiation, i.e. visible light, is composed of two parts; the real part ($\varepsilon_1$) and the imaginary ($\varepsilon_2$) part, which can be extracted from the measured MM data and the ellipsometric angles $\Psi$ and $\Delta$. Light propagation through a medium is typically described by the complex refractive index $N = n + i\kappa$, with  refractive index $n$ and  extinction coefficient $\kappa$ that is inseparably connected to the complex dielectric function by $\varepsilon = N^{2}$.  Hereby, the real part given by $\varepsilon_1 = n^2 - \kappa^2$ reflects mainly the ability of the material to polarize \cite{2007_Fujiwara}. The imaginary part of the dielectric function is associated with the losses, meaning the absorption of the electromagnetic field within the material described by $\varepsilon_2 = 2n\kappa$ \cite{2007_Fujiwara}.

Neither the experimentally determined MM elements $M_{ij}(E)$ nor $\Psi(E)$ and $\Delta(E)$ allows the direct determination of the desired dielectric tensor elements. By fitting the SIE spectra to an optical multilayer model, which accounts for the thickness and dielectric function of each layer, we can extract the dielectric functions and thickness information for the individual layers  \cite{1987_More, 1994_Drolet, 2012_Gilliot, 2020_Gu}. The fitting is achieved through a regression analysis that iteratively adjusts the model parameters until convergence is reached \cite{2007_Fujiwara, 2020_Gu}. Parameters that are well-known for individual layers can be fixed or used as starting points in the regression analysis, allowing for a more accurate and efficient extraction of the desired information. For the description of the CrSBr dispersion we use a multi- Lorentz oscillators approach.

Figure \ref{Figure_3} summarizes the measured and modeled Mueller-matrix elements $M_{ij}(E)$ normalized to the element $M_{11}$. The symbols describe the experiment while the lines are the modelled data as a result from the regression analysis. The model gives a good agreement with the measurement. 

\begin{figure}[h]
    \centering
    \includegraphics[width=0.45\textwidth]{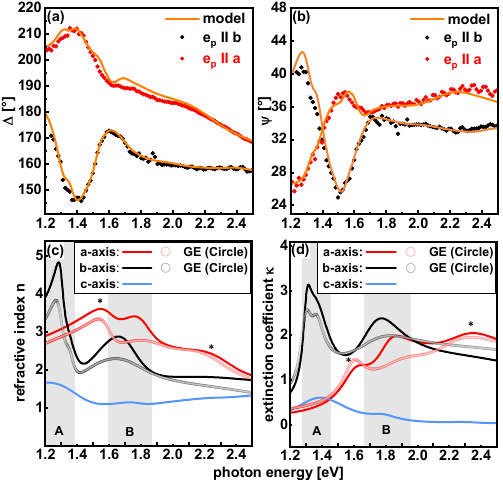}
    \caption{Generalized ellipsometry (GE) and Mueller-matrix analysis of CrSBr. (a, b) Ellipsometric spectra $\Delta(E)$ and $\Psi(E)$ for $\vec{e_p} \parallel a$  and $\vec{e_p} \parallel b$ and converged fit to the optical multilayer model (orange lines) [CrSBr on glass substrate]. (c, d) Extracted refractive index $n(E)$ and extinction coefficient $\kappa(E)$ for the principal optical axes $a$ (red), $b$ (black) and $c$ (blue), extracted from fits to the multilayer models by Mueller-matrix (solid lines) and GE (open circles) approach. Results from both models are in very good agreement. Shaded regions indicate the energy bands of the A- and B-exciton resonances centered around 1.3\,eV and 1.7\, eV, respectively, while asterisks denote additional polarization-dependent spectral resonances not yet assigned.}
\label{Figure_4}.
\end{figure}

The GE spectra $\Psi(E)$ and $\Delta(E)$ were measured with the samples axes $a$ (red) and $b$ (black) parallel to the plane of incidence as shown in Figure \ref{Figure_4}(a, b). The solid orange lines are the model result from the regression analysis to a suitable optical multilayer model as described above for the MM approach. The extracted refractive index $n(E)$ and extinction coefficient $\kappa(E)$ for the $a$-, $b$-, and $c$-axes (red/black/blue) obtained from the joint MM approach (solid lines) and GE approach (open circles) are provided in Figure \ref{Figure_4}(c, d). We decided to show the complex refractive index for easy comparison with absorption measurements and studies of polaritonic effects reported in the literature \cite{2022_Dirnberger,2023_Ruta, 2025_Ziegler}. Corresponding dielectric function $\varepsilon_{a,b,c}(E)$ are summarized in the SI.
In Figure \ref{Figure_4}(c, d), shaded bands mark the A- and B-exciton windows (centered around 1.3\,eV and 1.7\,eV); asterisks indicate additional, not yet assigned features. We find very good agreement for in-plane components of the complex refractive index $N_a$(E) and $N_b$(E) extracted from the MM and the GE approach.

We observe unique spectral signatures along all three crystallographic directions ($a$-, $b$-, and $c$-axes) as visible in the spectra displayed in Figure \ref{Figure_4}. Specifically, the $a$- and $b$-axis response show strong excitonic losses (absorption) in $\kappa (E)$, with the A exciton regime (~1.3\,eV ) being more pronounced along the $b$-direction. In the A-exciton band two strong features are visible. Those are assigned to the lowest energy and further excitonic band transitions. The secondary feature of about 60\,meV above the A-exciton can be tentatively interpreted as an excited excitonic state or a phonon sideband in agreement with strong polarization dependence \cite{2025_Smolenski}. In another less likely scenario the lower energy contribution could originate from charged excitons \cite{2025_Semina}. The strong absorption polarized along the $b$-axis is attributed to the strong electronic coupling arising from the quasi one-dimensional structure along the Cr-S chains \cite{2023_Klein}. The B-exciton regime ($\approx$1.7\,eV) exhibits a broader range of optical responses, with several excitonic resonances. Band-structure calculations show a strong contribution from the Brillouin-zone $X$-point to the B-exciton absorption \cite{2025_Liebich}. Unlike the A-exciton, the B-exciton band appears in both $a$ and $b$ axes, indicating it is not purely 1D but originates from multiple contributions with different band directions \cite{2025_Heißenbuettel, 2026_Markina, 2024_Komar}. Axis-dependent shifts of the resonance energies around the B-exciton band further support its multi-directional origin. Besides the main resonances which can be connected to A and B excitons we observe additional features of unknown origin in both in-plane polarizations (in Figure \ref{Figure_4} marked with $\ast$): one about 1.5\,eV, a second about 1.6\,eV and a third one about 2.7\,eV which potentially show additional exciton transitions which are also strongly polarization dependent. In contrast, the $c$-axis has a weak optical absorption consistent with its orientation perpendicular to the vdW layers of CrSBr which gives a reduced light-matter interaction \cite{2023_Klein,2025_Heißenbuettel,2023_Ruta}. Overall, the very good agreement of the model results from both measurement techniques for the in plane contribution demonstrate that GE is fully applicable given that the plane of incidence is properly aligned with respect to CrSBr's optical axes.

Our results confirm the highly anisotropic light-matter interaction of CrSBr. The determined optical anisotropy is consistent with quasi-1D  electronic states of CrSBr, which enforces direction-dependent dielectric behavior. Our SIE measurements performed on bulk CrSBr in paramagnetic state at room temperature reveal excitonic features that reflect the dependence on the direction of the electronic wavefunctions. Overall, the axes-dependent refractive indices and the absorption coefficients with $n_a \neq n_b \neq n_c$ and $\kappa_a \neq \kappa_b \neq \kappa_c$ extracted via SIE underscores the significance for understanding the complex light-matter interactions in the anisotropic 2D material CrSBr. We find reasonably good agreement between MM and GE results.The validity of GE measurements on CrSBr opens the door for cryogenic and magnetic field-dependent SIE measurements to study the dielectric tensor of CrSBr in the scientifically and technologically relevant antiferromagnetically and ferromagnetically ordered states.\\

The authors gratefully acknowledge financial support by the German Science Foundation (DFG) via the priority program 2244 (2DMP) through start-up funding. This project has received funding from the European Union’s Horizon Europe research and innovation program under grant agreement No. 101130224 “JOSEPHINE”.
Z.S. was supported by ERC-CZ program (project LL2101) from Ministry of Education Youth and Sports (MEYS), by project LUAUS25268 from Ministry of Education Youth and Sports (MEYS) and by the project Advanced Functional Nanorobots (reg. No. CZ.02.1.01/0.0/0.0/15-003/0000444 financed by the EFRR).

\section*{references}

%
% ****** End of file aipsamp.tex ******

\clearpage
\pagebreak

\begin{widetext}

\begin{center}
\vspace{4cm}
\textbf{\large Supplemental Information}\\
\vspace{1cm}
\textbf{\large Dielectric Tensor of CrSBr from Spectroscopic Imaging Ellipsometry}
\end{center}

\setcounter{equation}{0}
\setcounter{figure}{0}
\setcounter{table}{0}
\setcounter{page}{1}
\makeatletter
\renewcommand{\theequation}{S\arabic{equation}}
\renewcommand{\thefigure}{S\arabic{figure}}
\renewcommand{\bibnumfmt}[1]{[S#1]}
\renewcommand{\citenumfont}[1]{S#1}
\newenvironment{thebibliographySI}[1]
  {\begin{thebibliography}{#1}}
  {\end{thebibliography}}

\subsection*{Dielectric tensor for CrSBr}

\begin{figure}[h]
    \centering
    \includegraphics[width=1\linewidth]{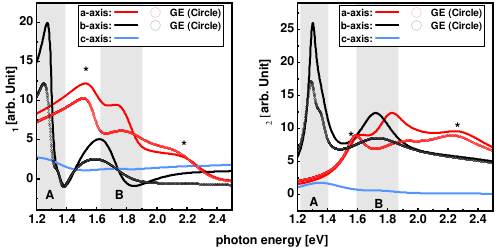}
    \caption{Real and imaginary parts of the dielectric function of CrSBr $\varepsilon_1$ and $\varepsilon_2$ at room temperature ($T = 300\,\mathrm{K}$) for the principal crystallographic axes $a$ (red), $b$ (black), and $c$ (blue). The solid lines correspond to the dielectric tensor components extracted from the Mueller-matrix analysis, whereas the open circles show the independently obtained in-plane GE results for the $a$- and $b$- axis. Dielectric functions extracted from both methods are in good agreement. The spectra reveal strong optical anisotropies: the A-exciton spectral range around $1.3\,\mathrm{eV}$ is most pronounced along the $b$-axis, whereas the broader B-excitonic band of resonances around $1.7\,\mathrm{eV}$ contributes to both in-plane axes. The dielectric response along the out-of-plane $c$-axis is considerably weaker. Asterisks ($\ast$) indicate additional polarization-dependent spectral features that are not assigned yet.}
    \label{fig:placeholder}
\end{figure}

\subsection*{Details of the generalized ellipsometry and Mueller-matrix measurements and modeling}

Spectroscopic imaging ellipsometry (SIE) was performed at room temperature ($T = 300\,\mathrm{K}$) in ambient atmosphere using an imaging ellipsometer in PCSA geometry (polarizer-compensator-sample-analyzer) \cite{1987_Azzam, 2007_Fujiwara,2010_Wurstbauer,2016_Funke}. All quantitative generalized ellipsometry (GE) and Mueller-matrix (MM) data were acquired in rotating-compensator ellipsometry (RCE) mode \cite{2022_Hilfiker,2007_Fujiwara}. In RCE mode, the compensator is rotated continuously and the detector signal is Fourier analyzed to extract either the ellipsometric angles $\Psi(E)$ and $\Delta(E)$ or a subset of normalized Mueller-matrix elements \cite{1987_Azzam,2007_Fujiwara,2022_Hilfiker}. For the PSCA geometry the first 3 rows of the Mueller-matrix of the sample are accessable. In order to measure the full Mueller-matrix, a second compensator after the sample needs to be part of the system. The spectra were recorded in the photon-energy range from \SI{1.2}{eV} to \SI{3.0}{eV} using a spectral step size of \SI{0.02}{eV} and \SI{0.05}{eV} for GE and MM respectively. The bandwidth of the grating monochromator depends on the wavelength, but can be assumed to be around \SI{5}{nm}.

For GE, CrSBr flakes on isotropic borosilicate glass (Borofloat) were measured at a fixed angle of incidence of $50^\circ$. The sample was manually rotated by $90^\circ$ between two measurements such that in one measurement configuration the plane of incidence was parallel to the crystallographic $a$-axis and in the other to the $b$-axis. CrSBr is strongly anisotropic (optical biaxial), with distinct principal optical axes along $a$, $b$, and $c$,\cite{2023_Klein,2024_Ziebel} and the crystallographic orientation was determined independently by polarization-resolved Raman spectroscopy and used for sample alignment \cite{2025_Mondal,2023_Pawake,2023_Torres}. In these two aligned geometries, the $p$-polarized component probes the optical response along the selected in-plane principal axis, while the orthogonal $s$-polarized component probes the perpendicular in-plane direction. Since the plane of incidence is aligned with the principal optical axes, additional cross-polarization generated by the CrSBr layer is minimized, should be zero for perfect alignmen. In such orientation the samples Jones matrix only has diagonal elements and the measured $\Psi(E)$ and $\Delta(E)$ are valid for chosen RCE algorithm \cite{1987_Azzam,2007_Fujiwara,2004_Schubert}.

For MM, CrSBr flakes on birefringent rutile (\ce{TiO2}) were measured over the same spectral range at angles of incidence of $50^\circ$ and $55^\circ$. In addition, the sample azimuth was varied using a rotation stage, and spectra were recorded for $\theta = 0^\circ$ and $75^\circ$. In the applied PCSA-RCE configuration, the experimentally accessible MM data correspond to the normalized elements of the first three rows of the Mueller matrix \cite{2022_Hilfiker,2021_Funke_Schiek}. Varying the sample azimuth changes the orientation of the biaxial crystal with respect to the laboratory frame, while the combination of different MM elements and angles of incidence provides sensitivity to polarization mixing and to the out-of-plane dielectric response \cite{2021_Funke_Schiek,2024_Schaper,2004_Schubert}. The Rutile substrate was characterized separately by using an uniaxial Cauchy model.

All spectra were analyzed using a regression-based optical multilayer model consisting of ambient / CrSBr / substrate \cite{2007_Fujiwara,2008_Hilfiker}. CrSBr was described throughout as a biaxial material with a diagonal dielectric tensor in its crystallographic frame,
\[
\boldsymbol{\varepsilon}(E)=\mathrm{diag}\!\left[\varepsilon_a(E),\varepsilon_b(E),\varepsilon_c(E)\right],
\]
which was transformed into the laboratory setup according to the sample azimuth and measurement geometry \cite{2004_Schubert,2021_Funke_Schiek,2024_Schaper}. For the GE data on Borofloat, the model was evaluated in the two aligned in-plane geometries and fitted within the same biaxial tensor description to $\Psi(E)$ and $\Delta(E)$ in order to determine the in-plane dielectric tensor components $\varepsilon_a(E)$ and $\varepsilon_b(E)$ \cite{1987_Azzam,2007_Fujiwara,2016_Funke}. For the MM data on rutile, all measured Mueller-matrix spectra at both angles of incidence and both azimuths were fitted simultaneously within one common biaxial tensor model in order to determine $\varepsilon_a(E)$, $\varepsilon_b(E)$, and $\varepsilon_c(E)$ consistently \cite{2021_Funke_Schiek,2022_Hilfiker,2024_Schaper}. The dielectric response of CrSBr was parameterized by Lorentz or Lorentz \cite{2007_Fujiwara,2008_Hilfiker}. In this way, GE provides an independent validation of the in-plane response in the aligned geometry, whereas the out-of-plane dielectric tensor component $\varepsilon_c(E)$ is constrained by the MM analysis which is mandatory for getting full access.

\subsubsection*{References}

\begin{thebibliographySI}{10}

\bibitem{1987_Azzam}
R.~M.~A Azzam and N.~M Bashara.
\newblock {\em Ellipsometry and Polarized Light}.
\newblock {North-Holland : Sole distributors for the USA and Canada, Elsevier
  Science Pub. Co.}, Amsterdam; New York, 1987.

\bibitem{2007_Fujiwara}
Hiroyuki Fujiwara.
\newblock {\em Spectroscopic Ellipsometry: Principles and Applications}.
\newblock John Wiley \& Sons, Chichester, 2007.

\bibitem{2010_Wurstbauer}
Ulrich Wurstbauer, Christian Röling, Ursula Wurstbauer, Werner Wegscheider,
  Matthias Vaupel, Peter~H. Thiesen, and Dieter Weiss.
\newblock Imaging ellipsometry of graphene.
\newblock {\em Applied Physics Letters}, 97(23):231901, 12 2010.

\bibitem{2016_Funke}
Funke, S, Miller, B, Parzinger, E, Thiesen, P, Holleitner, A~W, and
  U~Wurstbauer.
\newblock Imaging spectroscopic ellipsometry of mos2.
\newblock {\em Journal of Physics: Condensed Matter}, 28(38):385301, jul 2016.

\bibitem{2022_Hilfiker}
James~N. Hilfiker, Nina Hong, and Stefan Schoeche.
\newblock Mueller matrix spectroscopic ellipsometry.
\newblock {\em Advanced Optical Technologies}, 11(3-4):59--91, 2022.

\bibitem{2023_Klein}
Julian Klein, Benjamin Pingault, Matthias Florian, Marie-Christin
  Heißenb{\"u}ttel, Alexander Steinhoff, Zhigang Song, Kierstin Torres,
  Florian Dirnberger, Jonathan~B. Curtis, Mads Weile, Aubrey Penn, Thorsten
  Deilmann, Rami Dana, Rezlind Bushati, Jiamin Quan, Jan Luxa, Zdeněk Sofer,
  Andrea Alù, Vinod~M. Menon, Ursula Wurstbauer, Michael Rohlfing, Prineha
  Narang, Marko Lončar, and Frances~M. Ross.
\newblock The bulk van der waals layered magnet crsbr is a quasi-1d material.
\newblock {\em ACS Nano}, 17(6):5316--5328, 2023.
\newblock PMID: 36926838.

\bibitem{2024_Ziebel}
Michael~E. Ziebel, Margalit~L. Feuer, Jordan Cox, Xiaoyang Zhu, Cory~R. Dean,
  and Xavier Roy.
\newblock {{CrSBr}}: {{An Air-Stable}}, {{Two-Dimensional Magnetic
  Semiconductor}}.
\newblock {\em Nano Letters}, 24(15):4319--4329, April 2024.

\bibitem{2025_Mondal}
Priyanka Mondal, Daria~I. Markina, Lennard Hopf, Lukas Krelle, Sai Shradha,
  Julian Klein, Mikhail~M. Glazov, Iann Gerber, Kevin Hagmann, Regine~von
  Klitzing, Kseniia Mosina, Zdenek Sofer, and Bernhard Urbaszek.
\newblock Raman polarization switching in crsbr.
\newblock {\em npj 2D Materials and Applications}, 9(1):22, 03 2025.

\bibitem{2023_Pawake}
Amit Pawbake, Thomas Pelini, Nathan~P. Wilson, Kseniia Mosina, Zdenek Sofer,
  Rolf Heid, and Cl{\'e}ment Faugeras.
\newblock Raman scattering signatures of strong spin-phonon coupling in the
  bulk magnetic van der waals material {CrSBr}.
\newblock {\em Physical Review B}, 107(7):075421, February 2023.

\bibitem{2023_Torres}
Kierstin Torres, Agnieszka Kuc, Lorenzo Maschio, Thang Pham, Kate Reidy, Lukas
  Dekanovsky, Zdenek Sofer, Frances~M. Ross, and Julian Klein.
\newblock Probing defects and spin-phonon coupling in crsbr via resonant raman
  scattering.
\newblock {\em Advanced Functional Materials}, 33(12):2211366, 2023.

\bibitem{2004_Schubert}
Mathias Schubert.
\newblock {\em Infrared Ellipsometry on Semiconductor Layer Structures:
  Phonons, Plasmons, and Polaritons}.
\newblock Number v. 209 in Springer Tracts in Modern Physics. Springer, Berlin
  ; New York, 2004.

\bibitem{2021_Funke_Schiek}
Sebastian Funke, Matthias Duwe, Frank Balzer, Peter~H. Thiesen, Kurt Hingerl,
  and Manuela Schiek.
\newblock Determining the dielectric tensor of microtextured organic thin films
  by imaging mueller matrix ellipsometry.
\newblock {\em The Journal of Physical Chemistry Letters}, 12(12):3053--3058,
  2021.
\newblock PMID: 33739845.

\bibitem{2024_Schaper}
Sebastian Schaper~(né Funke), Matthias Duwe, and Wurstbauer Ursula.
\newblock Unambiguous determination of optical constants and thickness of
  ultrathin films by using optical anisotropic substrates.
\newblock (2502.21241v1), 2025.

\bibitem{2008_Hilfiker}
James~N. Hilfiker, Neha Singh, Tom Tiwald, Diana Convey, Steven~M. Smith,
  Jeffrey~H. Baker, and Harland~G. Tompkins.
\newblock Survey of methods to characterize thin absorbing films with
  {{Spectroscopic Ellipsometry}}.
\newblock {\em Thin Solid Films}, 516(22):7979--7989, September 2008.

\end{thebibliographySI}{10}
\end{widetext}

\end{document}